\documentclass[vecphys]{svmult}

\usepackage{makeidx}         
\usepackage{graphicx}        
\usepackage{multicol}        
\usepackage[bottom]{footmisc}

\makeindex             

\begin{document}

\title*{IMAGES: a unique view of the galaxy mass assembly since z=1}
\author{M. Puech\inst{1,2}, F. Hammer\inst{2}, H. Flores\inst{2}, Y.
  Yang\inst{2}, \and B. Neichel\inst{2}}
\institute{ESO, Karl-Schwarzschild-Strasse 2, D-85748 Garching bei M\"unchen, Germany
\texttt{mpuech@eso.org}
\and GEPI, Observatoire de Paris, CNRS, University Paris Diderot; 5
Place Jules Janssen, 92190 Meudon, France }
%
%
\maketitle

\begin{abstract} 
The Large Program IMAGES is near 2/3 of its completion. It provides us
with kinematics (GIRAFFE deployable IFUs), gas chemistry (FORS2),
detailed morphologies (HST/ACS) and IR photometry (Spitzer) for a set
of 70 galaxies representative of intermediate mass galaxies ($M_J$
$\leq$ -20.3 or 1.5$\times$10$^{10}$M$_\odot$) at z=0.4-0.75. We
discover that, 6 Gyr ago, a significant fraction of galaxies ($\geq$
40\%) had anomalous kinematics, i.e. kinematics significantly
discrepant from those of rotational or dispersion supported galaxies.
The anomalous kinematics cause the observed large dispersion of the
Tully-Fisher relation at large distances. IMAGES will soon allow us to
study distant galaxies at a level of detail almost comparable to that
of nearby galaxies.
\end{abstract}

\section{Introduction: intermediate-mass galaxies}
It is now relatively well-established that $\sim$50\% of the
present-day stellar mass has been formed since z=1. Most of this
stellar mass has been formed in intermediate-mass galaxies
(3$\!\times\!10^{10}$ to 3$\!\times\!10^{11}M_{\odot}$, i.e.,
$\sim$L$^*$ galaxies), as a result of strong star formation episodes
during which galaxies take the appearance of luminous infrared
galaxies (LIRGs, see \cite{hammer05}). However, it is still unclear
what physical processes have driven this evolution. To this respect,
internal kinematics of distant galaxies is a powerful tracer of the
major processes governing star-formation and galaxy evolution in the
early universe such as merging, accretion, and feedback related to
star-formation and AGN. Robustly measuring the internal kinematics of
distant galaxies is thus crucial for understanding how galaxies formed
and evolved.

\section{A representative sample of intermediate mass galaxies} 
A Large Program at VLT entitled IMAGES (Intermediate MAss Galaxy
Evolution Sequence) is aiming to derive both resolved kinematics and
integrated properties from VLT/GIRAFFE and FORS2 for galaxies selected
in the Chandra Deep Field (CDFS), and to combine these observations
with deep and high quality images from HST/ACS as well as with deep
mid-IR photometry from SPITZER/MIPS. This Large Program has now
reached about 2/3 of its completion, and recently, 39 additional
galaxies observed with GIRAFFE have been analyzed \cite{yang07}.
Combined with previous observations during the GTO \cite{flores06}, it
leads to a sample of 74 galaxies, which represents so far the largest
existing sample of resolved kinematics for distant galaxies.
Because in IMAGES we have deliberately selected
$M_J{\rm{(AB)}}$$<-20.3$ galaxies (i.e., with stellar masses larger
than 1.5$\times$10$^{10}$M$_\odot$), we assume a similar limit for the
combined sample. It let us with a sample of 63 galaxies which is well
representative of the luminosity function at z=0.4-0.75. Notice that
the combined sample includes galaxies from 4 independent fields of
view \cite{yang07} and is then unaffected by field-to-field variations
within Poisson statistics. Within this redshift range, GIRAFFE is able
to recover the kinematics of almost all galaxies with
$W_{0}(\mbox{{\sc[Oii]}})\ge15$\AA.


\section{Kinematics of distant galaxies}
At large distances the spatial resolution is not sufficient to resolve
the central regions of the galaxies. It implies that the observed
velocity dispersion ($\sigma$) of a rotational body is the convolution
of the actual random motions with the rotation. For a rotationally
supported galaxy, it unavoidably leads to a well defined peak in the
centre (see Fig. 1). We have developed a classification scheme which
allows us to compare any dispersion map to what it could be if it was
a rotational disk \cite{flores06}. Discrepancy from a rotational body
can be measured from differences in amplitude and in position of the
$\sigma$ peak between the pseudo rotational $\sigma$ map and the
observed one, which leads to a very robust diagnostic diagram for the
63 observed galaxies \cite{yang07}.

Among the 63 galaxies of the representative sample, we find 20
rotating disks (32\%), 16 rotating disks with perturbations (25\%) and
27 galaxies with complex kinematics (43\%, see Fig. 1 and
\cite{yang07}). Within this classification, perturbed rotations
correspond to a discrepant $\sigma$ peak from expectations for a pure
rotation. The complex kinematics class corresponds to objects for
which the large scale motions are not aligned to the optical major
axis, and show $\sigma$ map very discrepant from expectations for a
rotation (see Fig. 1).

\begin{figure}[!ht]
\begin{center}
\includegraphics[scale = 0.5]{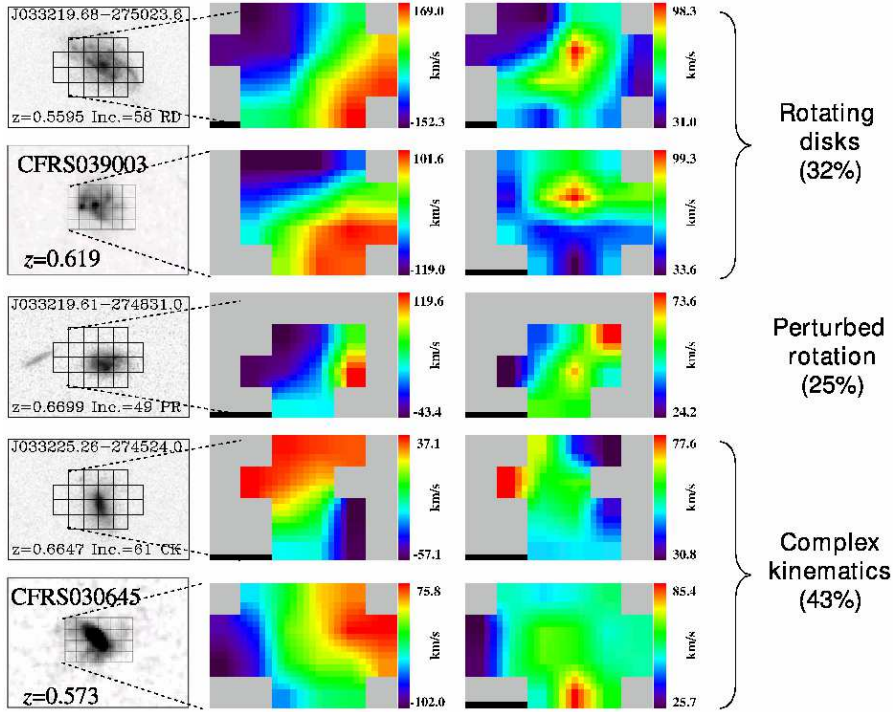}
\end{center}
\caption{Examples of kinematics of z=0.4-0.75 galaxies; Each row
  corresponds to one galaxy. From left to right: HST ACS F755W/F814W
  image, observed velocity field and $\sigma$-map. The two top rows
  show regular rotating disk; the three bottom rows show galaxies with
  anomalous kinematics, one with just a shift of the peak in the
  $\sigma$-map, e.g. a perturbed rotation, the two other with
  dynamical axis misaligned relatively to the main optical axis.}
\label{fig2}
\end{figure}

\section{Comparison with morphology}
Among these 63 galaxies, 52 have multi-band HST/ACS imaging. Neichel
et al. (in prep.) have constructed a new procedure to classify the
morphology of distant galaxies. This procedure relies on two important
ingredients which are the used of color maps (which helps to overcome
limitations due to k-morphological corrections) and the use of a
visual decision tree. This last ingredient is particularly important,
as they have shown that automatic methods can overestimate the
fraction of spirals by a factor of $\sim$2. They find a relatively
good agreement between the two classifications: most of Rotating Disks
are classified as Spirals, while most of galaxies having Complex
Kinematics show a peculiar morphology.

\begin{figure}[!ht]
\begin{center}
\includegraphics[scale = 0.5]{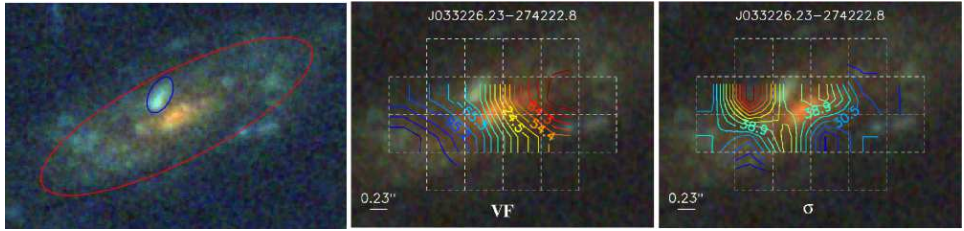}
\end{center}
\caption{Direct identification of a minor merger in a z$\sim$0.6
  galaxy classified as a perturbed rotator \cite{yang07}. \emph{Left:}
  3-band B-V-z HST/ACS image, with the blue ellipse indicating the
  position of the infalling satellite. \emph{Middle:} 3-band image
  with GIRAFFE isovelocities superimposed. The GIRAFFE IFU bundle is
  shown in white dashed-lines. \emph{Right:} 3-band image superimposed
  with GIRAFFE velocity dispersion map isocontours. The dispersion
  peak is shifted by one GIRAFFE pixel compared to the position of the
  stellar continuum of the satellite, which is due to shocks between
  the gas stripped out of the satellite during the interaction, and
  the gas of the main progenitor \cite{puech07}.}
\label{fig3}
\end{figure}

Combining morphology with spacially resolved kinematics allows us to
study distant galaxies in very fine details: for the first time, we
have recently detected a minor merger with a mass ratio of $\sim$1:18
in a z$\sim$0.6 galaxy classified as a perturbed rotator
\cite{puech07b} (see Fig. \ref{fig3}). Such processes, which are
predicted to be much more numerous than major mergers by numerical
models, could provide us with a very plausible mecanism for explaining
the kinematics of perturbed rotators.

\section{Dynamical state of distant galaxies} 
Let us now consider a representative sample of intermediate mass
galaxies: at z=0.6, 60\% of galaxies have $W_{0}([OII])\ge15$\AA
\cite{hammer97}. Let us assume that quiescent ($W_{0}([OII])<15$\AA)
galaxy are either rotationally or dispersion supported, thus
minimising the fraction of galaxies with anomalous kinematics. Then
our results imply that 42$\,\pm\,$7\% of z=0.4-0.75 galaxies have
anomalous kinematics, including 26$\,\pm\,$7\% possessing complex
kinematics. Because up to 97\% of local intermediate mass galaxies are
either E, S0 or spirals \cite{hammer05}, it is likely that the
fraction of anomalous kinematics is close to a few percents today.
This leads to an extremely rapid evolution of kinematical properties
of galaxies, with about 10 times more complex kinematics about 6 Gyr
ago.

The observed evolution of the Tully-Fisher Relation (TFR) provides us
with a strong and independent confirmation. It has been suggested that
galaxies with non-relaxed kinematics (PR and CK) are responsible for
the very large dispersions of the TFR at high redshift
\cite{flores06}. This result is confirmed by Puech et al. (in prep)
using the new sample of 63 galaxies: all the dispersion of the distant
TFR can be accounted for by galaxies having non-relaxed kinematics. So
it is beyond doubt that kinematics is among the most rapidly evolving
properties of galaxies.

Which physical process could explain such a dramatic evolution ?
Anomalous kinematics are linked with strong variations of the specific
angular momentum consistent with a random walk evolution due to
merging between galaxies, as predicted by the hierarchical scenario of
galaxy formation \cite{puech07}. Indeed, during a merger, and
especially a major merger, galaxies pass through various stages during
which the disk may be destroyed, generating significant discrepancies
to the general behaviour of isolated rotating disks.

\section{Conclusion} 
We do find a strong evolution of the galaxy kinematics since z=0.6,
with a significant fraction of galaxies with complex kinematics.
Observations presented here tell us that major mergers could have
played an important role in shaping galaxies as we observe them today.
Indeed many estimates of the merger rate found that a typical
intermediate mass galaxy should have experienced 0.5 to 0.75 major
merger since z= 1 (see \cite{rawat07,kartaltepe07} and references
therein), and most of them since z=2-3. A scenario in which merging is
the dominant physical process explains many evolutionary features
since z=1, including the number density evolution of LIRGs and the
emergence of galaxies with complex morphologies and with blue cores at
z$>$ 0.4 \cite{hammer05}. It is moreover particularly in agreement
with the fact that the Milky Way has had an exceptionally quiet merger
history \cite{hammer07}. Present-day $M*$ galaxies being mostly spiral
galaxies, it is unlikely that they have all escaped a major merging
since z=2-3.

%
%
%


\printindex
\end{document}